\documentclass[10pt]{article}

 \usepackage{amsmath,amsthm,amssymb}
 \usepackage{color,colortbl}
 \usepackage{fancyhdr}
 \usepackage{xcolor,pict2e}
 \usepackage{graphicx}
 \usepackage{cite}
 \renewcommand{\headrulewidth}{0pt}

 \definecolor{myaqua}{rgb}{0.0,0.5,0.55}
 \definecolor{lightaqua}{rgb}{0.75,0.95,0.95}

\usepackage{caption}
\usepackage{floatrow}

\def\lin#1#2{\textcolor[rgb]{0.6,0.6,0.6}{\vspace*{#1mm} \hrule
   height 3 pt \vspace*{#2mm}}}
\def\bt{\begin{tabular}}
\def\et{\end{tabular}}
\def\and{\mbox{ and }}

\def\1{{\bf 1}}

 \def\sectionn#1{\refstepcounter{section}{\color{myaqua}

 \vskip 6mm

 \noindent\Large\bf\thesection. #1}

 \vskip 3mm}

\renewcommand{\headrulewidth}{0.5pt}

\renewcommand{\headrule}{\hbox to\headwidth{\color{myaqua}\leaders\hrule height \headrulewidth\hfill}}

\begin{document}

 $\mbox{ }$

 \vskip 12mm

{ 

{\noindent{\huge\bf\color{myaqua}
  Arguing on entropic and enthalpic first-order phase transitions in strongly interacting matter
  }}
\\[6mm]
{\large\bf F. Wunderlich$^{1,2}$, R. Yaresko$^{1,2}$, B. K\"ampfer$^{1,2}$}}
\\[2mm]
{
 $^1$Helmholtz-Zentrum Dresden-Rossendorf, Inst. f\"ur Strahlenphysik,
PF 510119, D-01314 Dresden, Germany\\
$^2$Institut f\"ur Theoretische Physik, TU Dresden, D-01062 Dresden, Germany}

\includegraphics{pic2.ps}

\lin{5}{7}

 {\noindent{\large\bf\color{myaqua} Abstract}{\bf \\[3mm]
 \textup{
 The pattern of isentropes in the vicinity of a first-order phase transition is proposed as a
 key for a sub-classification. While the confinement--deconfinement transition, conjectured 
 to set in beyond a critical end point in the QCD phase diagram, is often related to an entropic
 transition and the apparently settled gas-liquid transition in nuclear matter is an enthalphic transition, 
 the conceivable local isentropes w.r.t.\ "incoming" or "outgoing" serve as another useful guide
 for discussing possible implications, both in the presumed hydrodynamical expansion stage
 of heavy-ion collisions and the core-collapse of supernova explosions. Examples, such as the
 quark-meson model and two-phase models, are shown to distinguish concisely the different transitions.   
 }}}
 \\[4mm]
 {\noindent{\large\bf\color{myaqua} Keywords}{\bf \\[3mm]
 Entropic and enthalpic phase transitions, chiral phase transition, isentropes, 
 quark-meson model, linear sigma model with linearized fluctuations
}}

\lin{3}{1}

\sectionn{Introduction}

{ \fontfamily{cmr}\selectfont
 \noindent 
The beam energy scan at RHIC \cite{Adare:2012uk,Adare:2012wf,Adamczyk:2013gw,Adamczyk:2014mzf,Soltz:2014dja,Das:2014oca,McDonald:2015tza}
is aimed at searching for a critical end point (CEP) in
the phase diagram of strongly interacting matter, which is related to confinement-deconfinement effects. 
At a CEP \cite{Rischke:2003mt,Stephanov:2004wx,Fukushima:2010bq,Friman:2011zz}, 
a line of a first-order
phase transitions (FOPT) is conjectured to set in. Still, the hypothetical CEP could not (yet) be
localized by ab initio QCD calculations. Therefore, details of the FOPT curve and details of
the equation of state in its vicinity are unsettled to a large extent. 

The utmost importance of the search for a CEP is also manifested by the fact that further ongoing
relativistic heavy-ion collision experiments, such as NA61/SHINE 
\cite{Gazdzicki:2008pu,Czopowicz:2015mfa,Aduszkiewicz:2015jna,MajaMackowiak-PawlowskafortheNA61/SHINE:2016vzn}, 
have it on the their priority list, and planned 
experiments at FAIR, e.g.\ CBM \cite{Chattopadhyay:2014mha}, at NICA, e.g.\ by the MPD group \cite{Kekelidze:2015yza},
and at J-PARC, e.g.\ by the J-PARC heavy-ion collaboration \cite{Sako:2015cqa}, are primarily motivated by it.
The proceedings of the CPOD conferences \cite{LawrenceBerkeleyLab.:2013uoa,Proceedings:2015exj} 
document well the theoretical expectations
and experimental achievements in this field.

The CEP itself (which may occur also as a tricritical point \cite{Ding:2015ona}) is interesting, as it is expected 
to show up in specific fluctuation observables
\cite{Stephanov:1999zu,Gupta:2009mu,Mohanty:2009vb,Adamczyk:2013dal,Adamczyk:2014fia,Almasi:2016hhx}, 
related to critical exponents, however, 
also the emerging FOPT curve can give rise to interesting
physics phenomena.  If the hypothetical FOPT curve continues to small or even zero 
temperatures, astrophysical consequences for neutron stars 
\cite{Kampfer_neutronstars,Schertler:1997za,Schertler:1999xn,Schertler:2000xq,Macher:2004vw,
Dexheimer:2008ax,Pagliara:2009dg,Kurkela:2010yk,Fischer:2010wp,Alford:2013aca,
Yasutake:2014oxa,Zacchi:2015lwa,Hempel:2015vlg,Alvarez-Castillo:2015xfa}
proto-neutron star formation and core-collapse supernova explosions 
\cite{Dexheimer:2008ax,Sagert:2008ka,Fischer:2010wp,Nishimura:2011yb,Pan:2015sga}
are directly related to the physics of heavy-ion collisions, supposed the FOPT curve is
accessible in such experiments (cf. \cite{Bugaev:2014bua} for searches for two-phase mixture effects
related to the deconfinement FOPT). 

From the theory side, the famous Columbia plot (cf.\
\cite{Ding:2015ona}
for an update) unravels the following qualitative features:
(i) At zero chemical potential, three-flavor QCD in the chiral limit displays
a first-order confinement-deconfinement transition which extends to non-zero
strange-quark masses $m_s < m_s^\text{tri}$ and light-quark masses
$m_{u,d} \to 0$; the delineation curve to the region $m_{u,d,s,} > 0$
is related to a 2nd order transition with $Z(2)$ symmetry, beyond which
the transition turns into a cross over; for $m_s > m_s^\text{tri}$ and
$m_{u,d} \to 0$, the 2nd order transition line is related to $O(4)$ symmetry.
The physical point $m_{u,d,s} > 0$ is in the cross over region.
(ii) For $m_s > m_s^\text{tri}$ and $m_{u,d} \to 0$, the phase structure
in the temperature-chemical potential plane is determined by a 2nd order
transition curve of presumably negative slope (with the above mentioned
universal $O(4)$ scaling properties) which ends in a tri-critical point, where the
1st order transition sets in, expected to continue to zero temperature.
(iii) Upon enlarging $m_{u,d}$ toward the physical values
and keeping the conjectured $m_s > m_s^\text{tri} > 0$, the 2nd oder transition
curve turns into the pseudo-critical (cross over) curve which ends at
non-zero chemical potential in a CEP. The latter one can be thought to arise
from the previous tri-critical point along a 2nd order $Z(2)$ curve when enlarging
$m_{u,d}$. Therefore, the expectation for 2+1 flavor QCD with physical
quark masses is the existence of a CEP at a temperature below the
pseudo-critical temperature of $(154 \pm 9)$~MeV and non-zero chemical
potential and an emerging 1st oder transition curve going to zero temperature
\cite{Ding:2015ona}.
Present day lattice QCD evaluations attempt to quantify these features,
cf.\ \cite{Bellwied:2015rza}, for example.

In a recent series of papers \cite{Hempel:2013tfa,Steinheimer:2013xxa,Iosilevskiy:2015sia},
the authors promote
a useful sub-classification of FOPTs by attributing the confinement--deconfinement transition
to an entropic one, while the established gas-liquid transition in nuclear matter \cite{Rischke:2003mt,Stephanov:2004wx,Fukushima:2010bq,Friman:2011zz}
is classified as enthalpic one. The key is the Clausius-Clapeyron equation
\begin{equation}
\frac{d p_c (T)}{d T}  = \frac{s_1 /n_1 - s_2 / n_2}{1/n_1 - 1/n_2} \label{clausius_1}
\end{equation}
which relates the slope of the critical pressure, $p_c$, along the FOPT w.r.t.\
temperature, $T$, to entropy densities $s_{1,2}$ and baryon densities $n_{1,2}$.
Denoting by the label "1" the dilute (confined/hadron) phase 
and by "2" the dense (deconfined/quark-gluon) phase, the slope 
of the critical pressure curve is positive, $d p_c / dT > 0$, for larger entropy per baryon
in phase "1", meaning an enthalpic FOPT. In contrast, for larger entropy per baryon in phase "2" the critical
curve has a negative slope, $d p_c / dT < 0$ meaning an entropic FOPT.

Some guidance for the trajectories of fluid elements is given by the isentropic curves,
determined by $s/n = const$, when having in mind the adiabatic expansion of matter created
in the course of a heavy-ion collision as long as the respective fluid element is in a pure
phase, "2" or "1". The details of the transit through the two-phase coexistence region depend
on the latent heat and other details of the equation of state. 
With respect to investigations of the heavy-ion dynamics (cf.\ \cite{Steinheimer:2007iy})
seeking for imprints of the conjectured QCD FOPT and CEP signatures, it seems tempting
to  clarify in a clear-cut picture the different patterns of isentropes being related to a FOPT.  
  
Our note is organized as follows. In section \ref{sec:patterns} we discuss obvious types of isentropic 
patterns which may accompany a FOPT in strongly interacting matter. The pattern
classification is put in relation to the entropic and enthalpic sub-classes. We see enthalpic
transitions either with incoming-only or incoming+outgoing isentropes, thus
qualifying also the latter one for modeling the QCD deconfinement-confinement  
transition. Examples based on transparent models are presented in section \ref{sec:examples} and appendix \ref{apdx:HQ}.
In section \ref{sec:summary}, we summarize.
}
\sectionn{Isentropic patterns}
\label{sec:patterns}
{ \fontfamily{cmr}\selectfont
 \noindent

We restrict our discussion to the grand canonical description of matter by an equation
of state $p(T, \mu)$ with one conserved charge, e.g.\ baryon number, related to the chemical
potential $\mu$. Entropy density and baryon density are given by 
$s(T, \mu) = \partial p / \partial T$ and $n(T, \mu) = \partial p / \partial \mu$ 
and the Gibbs -Duhem relation $e + p = sT + n \mu$ holds ($e$ is the energy density). 
Considering the
region $s > 0$ and $n > 0$, the isobars $ p = const$ have negative slopes in the \mbox{$T$--$\mu$}
diagram upon $d T / d \mu \vert_{dp = 0} = - n/s$.
We assume locally a FOPT which is signaled by a kinky behavior of $p(T, \,u)$ over the
\mbox{$T$--$\mu$} plane, both in $T$ and $\mu$ directions. $p(T, \mu)$ refers here to stable states;
if multi-valued regions emerge, the branch with maximum pressure is the stable one.
We further assume, for the sake of definiteness, the FOPT curve has  a negative slope, 
$d T_c (\mu) / d \mu < 0$.
In fact, $d (p_1(T, \mu) - p_2(T, \mu) = 0$ on the FOPT curve delivers
$d T_c / d \mu = -(n_1 - n_2)/(s_1 - s_2)$, where we suppose $n_1 < n_2$ and
$s_1 < s_2$. 

We also recall from the equilibrium conditions $T_1 = T_2$, $\mu_1 = \mu_2$
and $p_1 = p_2$ on the FOPT curve the relation
\begin{equation}
\frac{d p_c(\mu) }{d \mu} = \frac{n_1 /s_1 - n_2/s_2}{1/s_1 -1/s_2} \label{clausius_2}
\end{equation} 
which is another form of the Clausius-Clapeyron equation \eqref{clausius_1}.

\begin{figure}[ht]
\begin{tabular}{rccc}
&type IA & type IB & type II\\
\put(0,40){$T$}
\put(370,-15){$\mu$}
\put(10,-5){\thicklines\vector(0,1){70}}
\put(10,-5){\thicklines\line(1,0){125}}

\put(135.5,-5){\thicklines\line( 1, 2){3}}
\put(135.5,-5){\thicklines\line(-1,-2){3}}
\put(138.5,-5){\thicklines\line( 1, 2){3}}
\put(138.5,-5){\thicklines\line(-1,-2){3}}

\put(138.5,-5){\thicklines\line(1,0){126.5}}

\put(265.5,-5){\thicklines\line( 1, 2){3}}
\put(265.5,-5){\thicklines\line(-1,-2){3}}
\put(268.5,-5){\thicklines\line( 1, 2){3}}
\put(268.5,-5){\thicklines\line(-1,-2){3}}

\put(268.5,-5){\thicklines\vector(1,0){130}}&

\includegraphics[clip=true, trim= 0mm 20mm 0mm 20mm, height=0.15\textwidth]{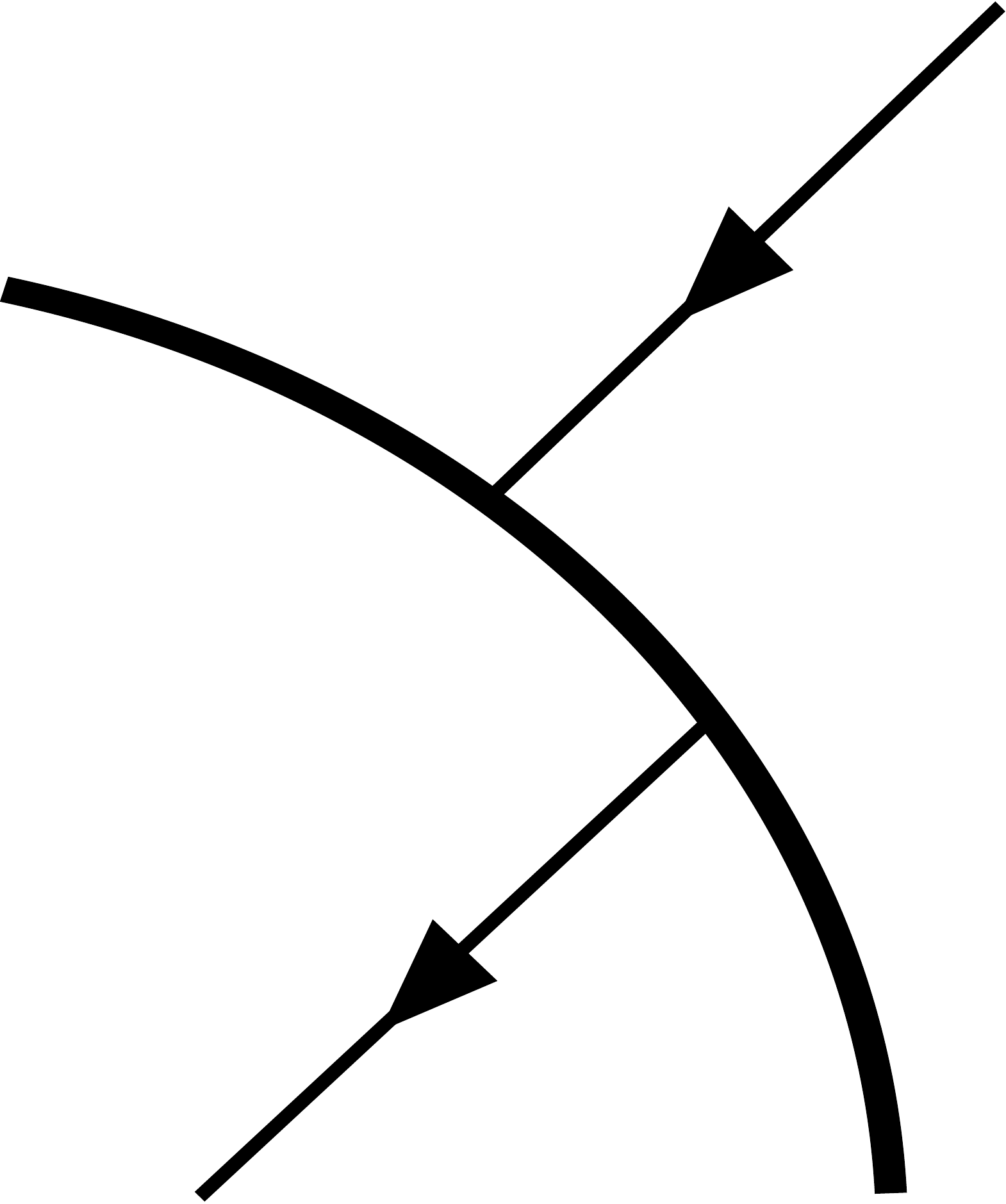} &
\includegraphics[height=0.15\textwidth]{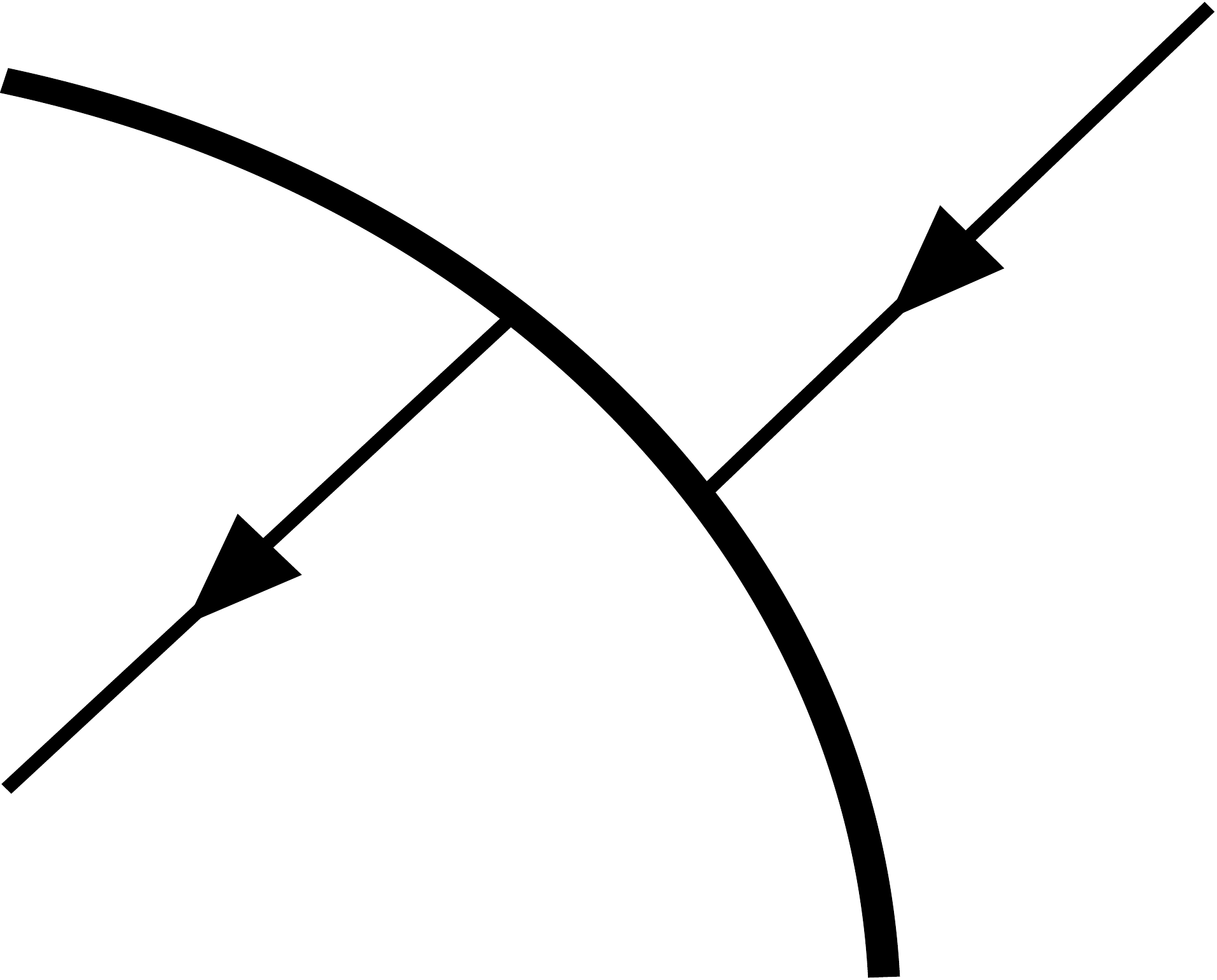} & 
\includegraphics[height=0.15\textwidth]{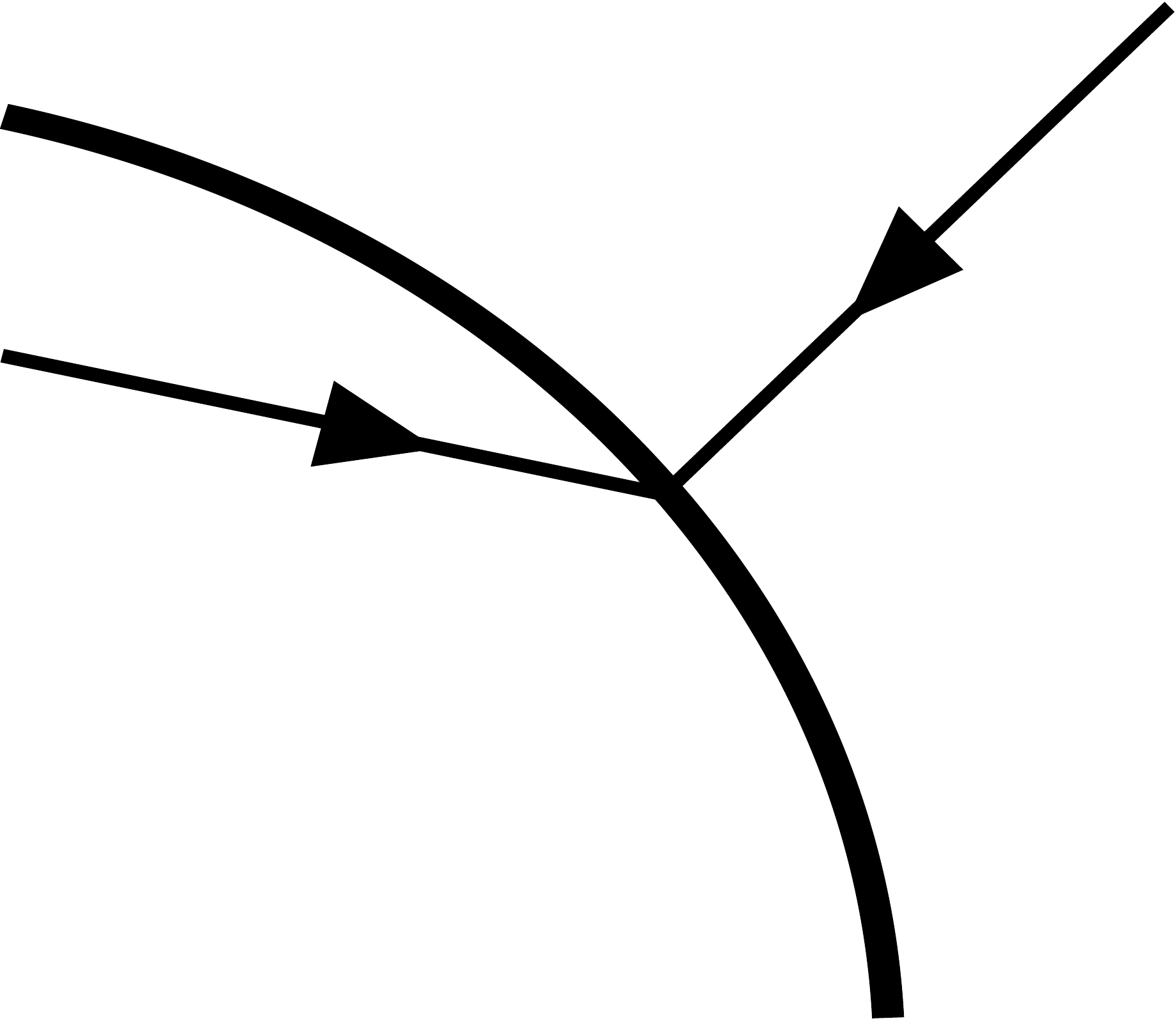} \\[2em]

\put(0,40){$T$}
\put(370,-15){$n$}
\put(10,-5){\thicklines\vector(0,1){70}}
\put(10,-5){\thicklines\line(1,0){125}}

\put(135.5,-5){\thicklines\line( 1, 2){3}}
\put(135.5,-5){\thicklines\line(-1,-2){3}}
\put(138.5,-5){\thicklines\line( 1,2 ){3}}
\put(138.5,-5){\thicklines\line(-1,-2){3}}

\put(138.5,-5){\thicklines\line(1,0){126.5}}

\put(265.5,-5){\thicklines\line( 1, 2){3}}
\put(265.5,-5){\thicklines\line(-1,-2){3}}
\put(268.5,-5){\thicklines\line( 1, 2){3}}
\put(268.5,-5){\thicklines\line(-1,-2){3}}

\put(268.5,-5){\thicklines\vector(1,0){130}}&

\includegraphics[width=0.25\textwidth]{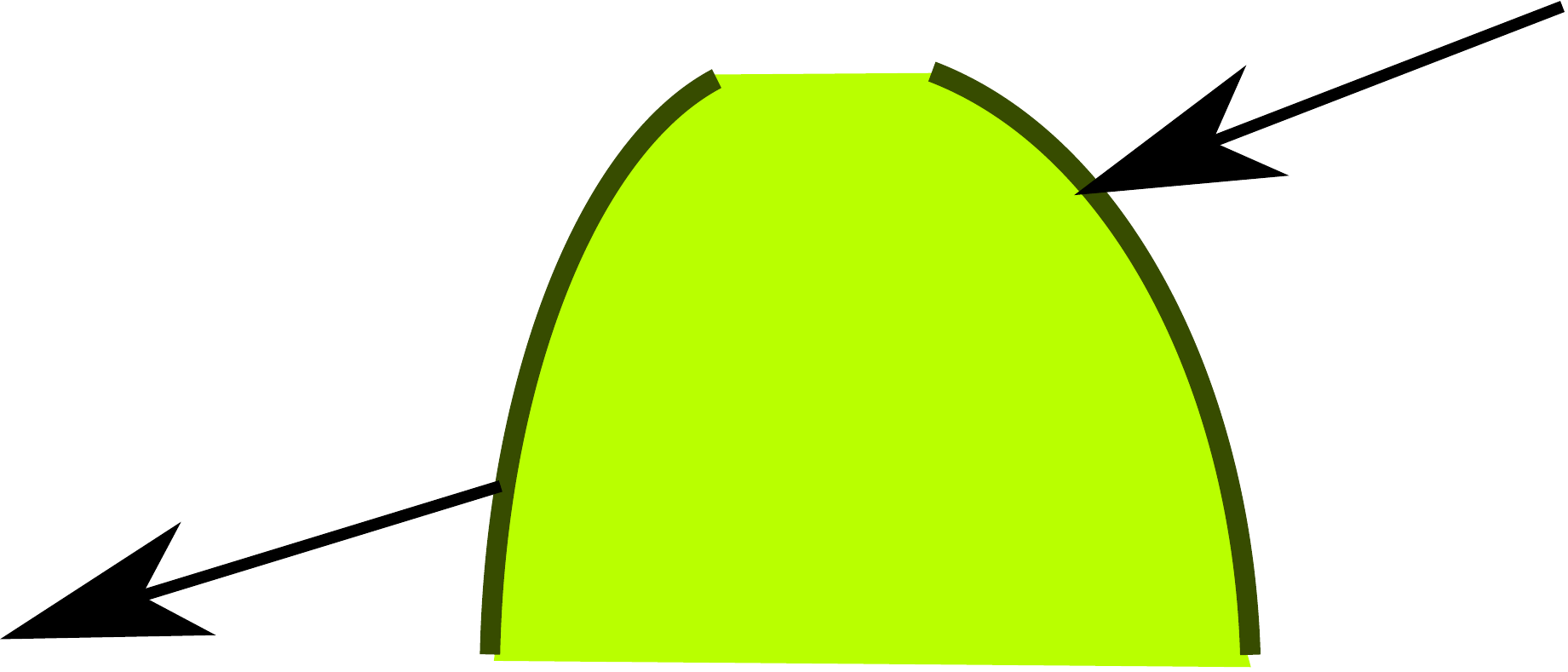} &
\includegraphics[width=0.25\textwidth]{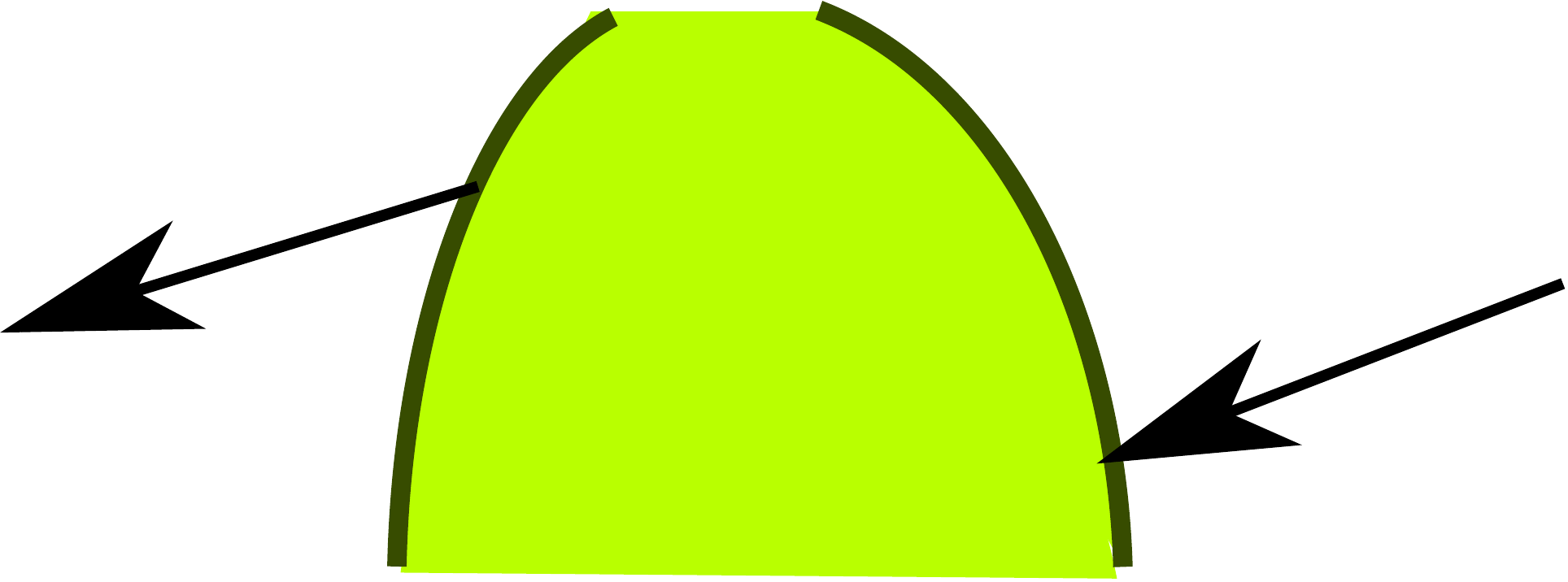} &
\hspace{2mm}\includegraphics[clip=true, trim= 0mm 0mm 0mm 10mm, width=0.25\textwidth]{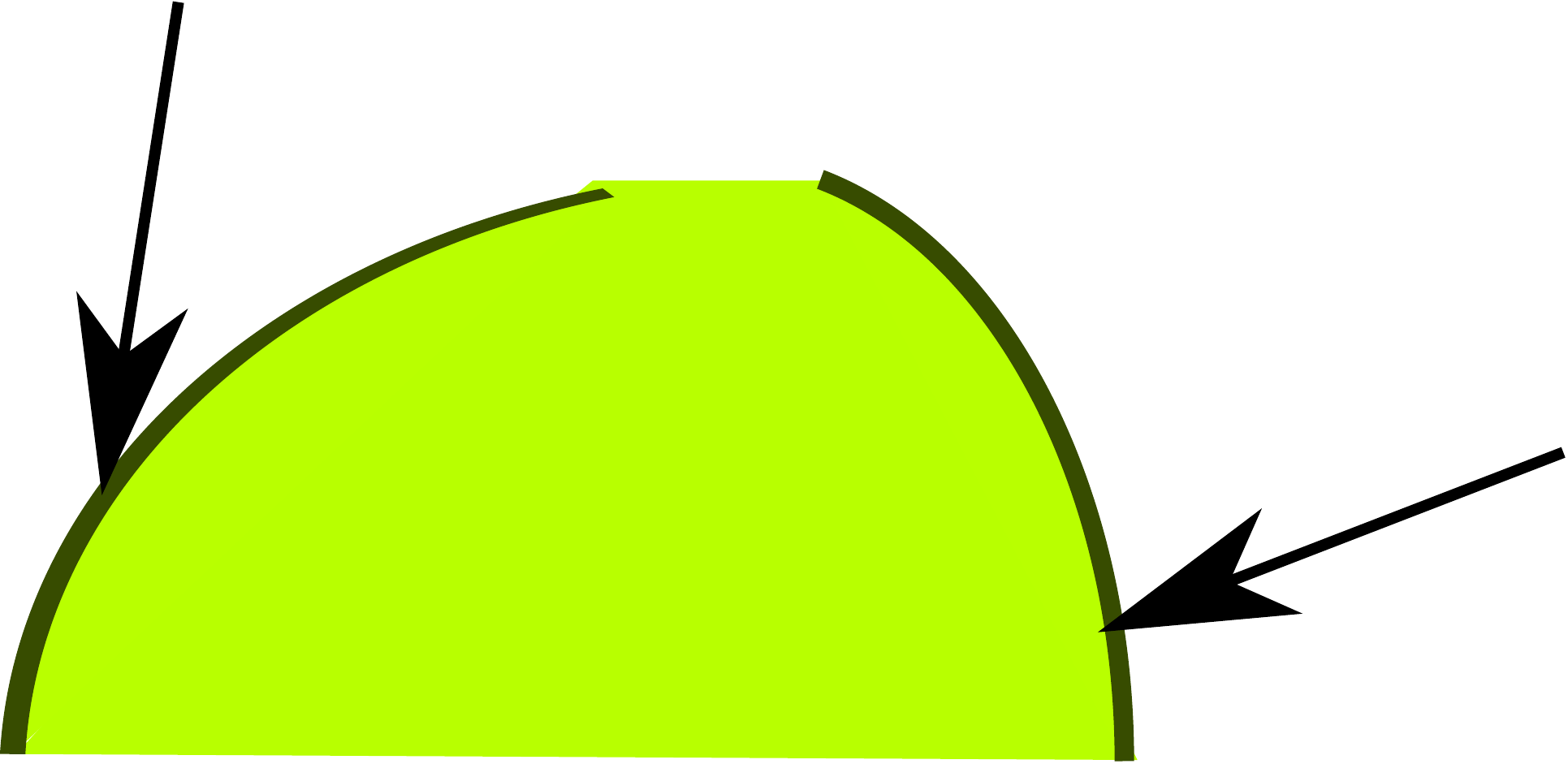}  
\end{tabular}


\caption{Schematic representation of isentropes (lines with arrows indicating the expansion path)
         for the FOPT types IA (left panels, $s/n =const$), IB (middle panels, $s/n = const$) 
         and Typ II (right panels, $s_1/n_1 > s_2/n_2$) in the \mbox{$T$--$\mu$} plane (upper row) 
         and the \mbox{$T$--$n$} plane (lower row). States in "1" (see text) are left/below the phase 
         border line (fat curves in the upper row), 
         while states in "2" are right/above. The green areas in the lower row depict a part of the 
         two-phase coexistence regions for the respective types.
         Note that the coexistence regions (green areas) can appear in quite different shapes.
\label{fig:0}}
\end{figure}
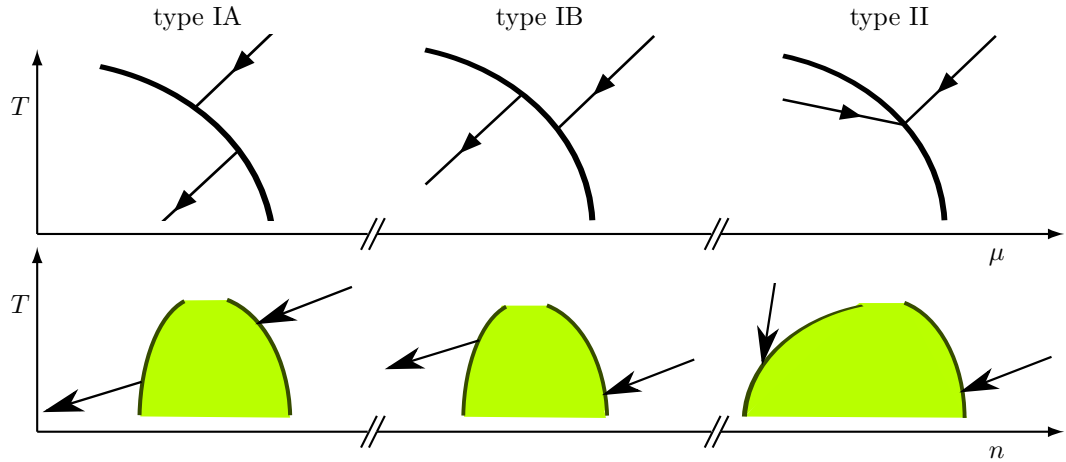

From selected examples we can infer three different patterns of isentropes in the
\mbox{$T$--$\mu$} plane:\\
Type IA: Isentropes come in from the phase "2", enter the critical curve $T_c(\mu)$ and leave
it toward the phase "1" at {\sl lower} temperature, see Fig.~\ref{fig:0}, left top panel.
According to Clausius-Clapeyron \eqref{clausius_1} one has $d p_c(T) / dT > 0$, i.e.\
a gas-liquid or enthalpic transition in the nomenclature of \cite{Iosilevskiy:2015sia}.\\
Type IB: Isentropes come in from the phase "2", enter the critical curve $T_c(\mu)$ and 
evolve toward phase "1" at {\sl higher} temperature, see Fig.~\ref{fig:0} middle top panel.
Clausius-Clapeyron tells us for that case $dp_c (T) / dT < 0$, i.e.\ a QCD type or
entropic FOPT in the nomenclature of \cite{Iosilevskiy:2015sia}.\\
Type II: Isentropes come in from both sides, i.e.\ phases "1" and "2", enter the critical curve
$T_c (\mu)$ and run down on it, see Fig.~\ref{fig:0}, right top panel.
According to our experience with a number of models, $s_2/n_2 < s_1/n_1$ in a point
on the critical curve, i.e.\ also a gas-liquid type or enthalpic FOPT with
$dp_c(T) /dT > 0$.

The direction of isentropes is such to describe expansion, i.e.\ both temperature and density
drop in pure phases. Type I is related to in-out (or going-through) isentropes, while type II has incoming-only.
A prominent example for type II is the van der Waals equation of state, cf.\ \cite{johnston2014} and 
figure 1 in \cite{yuen2015}. We emphasize the local character of our
consideration, that is the restriction to the vicinity of a \mbox{$T$--$\mu$} point on the presumed
phase boundary. These patterns translate directly into the \mbox{$T$--$n$} plane, see bottom row
of Fig.~\ref{fig:0}, where one verifies that dropping temperatures along isentropes in pure phases
imply in fact dropping densities too, i.e. proper expansion. Types IA and IB are delineated
by $s_1/n_1 = s_2/n_2$, resulting in $p_c(T) = const$.
Types IA and II share as common
feature flatter isobars than the critical curve $T_c(\mu)$; for type IB, the critical curve is
flatter than the isobars. 

For the moment being we do not see the need to study further fine details, e.g.\ slopes
and relative slopes of isentropes near the critical curve.

We would like to emphasize that also models of type IA could serve as an illustration of the
possible structure of the phase diagram, despite they belong to the gas-liquid transition
type: Suppose $n_1^c (T \to 0) > n_0$, where $n_0$ is the nuclear saturation density and
$n_1^c$ denotes the density of phase "1" at the critical curve, then nothing seems to speak
against the scenario with an expanding and cooling fluid element initially in phase "2", which
traverses the confinement transition region (two-phase coexistence) and arrives in the 
hadronic world of phase "1". 
That means, if "2" is a deconfined state, then both IA and IB allow for a graceful exit into the
pure (hadronic) phase "1", while II ends locally in a two-phase mixture of "1+2" for adiabatic
expansion dynamics, i.e. some part of matter remains in the deconfined state "2", e.g.\
as quark nuggets, contrary to our present expectations and in agreement with the failure of previous searches 
for them \cite{Schaffner:1991qg,Schaffner:1993nn,Adams:2005cu,Madsen:2006yw,Adriani:2015epa}, 
(see however \cite{Gorham:2012hy,Atreya:2014sca} for considering them as candidates of dark matter).   
Whether realistic models can be designed to do so (cf. \cite{Benic:2015pia} for a recent attempt), in agreement
with serving for two-solar mass neutron stars, is a question beyond the schematic phenomenological 
approaches. 
Anyhow, type IA supplements the considerations favored in 
\cite{Iosilevskiy:2015sia,Steinheimer:2013xxa}.
}

\sectionn{Examples}
\label{sec:examples}

{ \fontfamily{cmr}\selectfont
 \noindent

We are going to present a few examples for the above discussed transition types. For that,
we select the quark-meson model\footnote{
We chose this since in the chiral limit it obeys the same symmetries 
(an $O(4)\simeq SU(2)\times SU(2)$\cite{Tetradis:2003qa}) as 
QCD \cite{Pisarski:1983ms} putting both into the same universality class 
and thus rendering the model a good prototype for studying
the properties of the QCD chiral transition} (cf.\ \cite{Wunderlich:2015rwa} for a description of the setting used
here\footnote{
In a nutshell, the employed model, also coined linear sigma model, is based on a doublett of quark degrees
of freedom, an iso-scalar sigma field and an iso-triplett pion field with standard coupling among these fields.})
with linearized meson field fluctuations\footnote{
According to our experience with numerical evaluations, the account of linearized meson field fluctuations
modifies significantly the results of the mean field approximation. (For the inclusion of the complete fluctuations 
spectrum within the functional renormalization group approach, see \cite{Tripolt:2013jra}.) In particular, the 
fluctuating meson degrees of freedom deliver explicit contributions to the pressure}
and show that only shifting the nucleon/quark vacuum mass parameter
$m_{q,ß}$ relative to the critical chemical potential at zero temperature $\mu_c^0$ is sufficient to switch from IA to II.
The latter one is to a large extent determined by the product of the sigma mass parameter $m_\sigma$ and the 
(classical) vacuum expectation value of the sigma field $\langle \sigma\rangle_0$. 
We are fully aware of the shortcomings
of such a model w.r.t.\ proper account of nuclear matter properties at low temperatures
and QCD thermodynamics at high temperatures, as discussed in \cite{Steinheimer:2013xxa}. 
But in view of the pertinent complexity
of the QCD degrees of freedom in the strong coupling regime such a model with chiral
symmetry breaking and restoration may give some glimpses of what is conceivable, in 
principle. 

Also our model for the type IB (cf. appendix \ref{apdx:HQ}) has, at best, illustrative character: It is a two-phase construction
with states in "2" modeled by the extrapolation of weakly interacting quarks and gluons, 
supplemented by an effective bag constant to account for some non-perturbative aspects,
and states in "1" referring to thermal light-meson (pion) excitations and nucleons in some
mean field approximation including a realistic incompressibility modulus.  
 
\begin{figure}
{\includegraphics[clip=true,trim=7mm 1mm 15mm 13mm,width=0.48\textwidth]{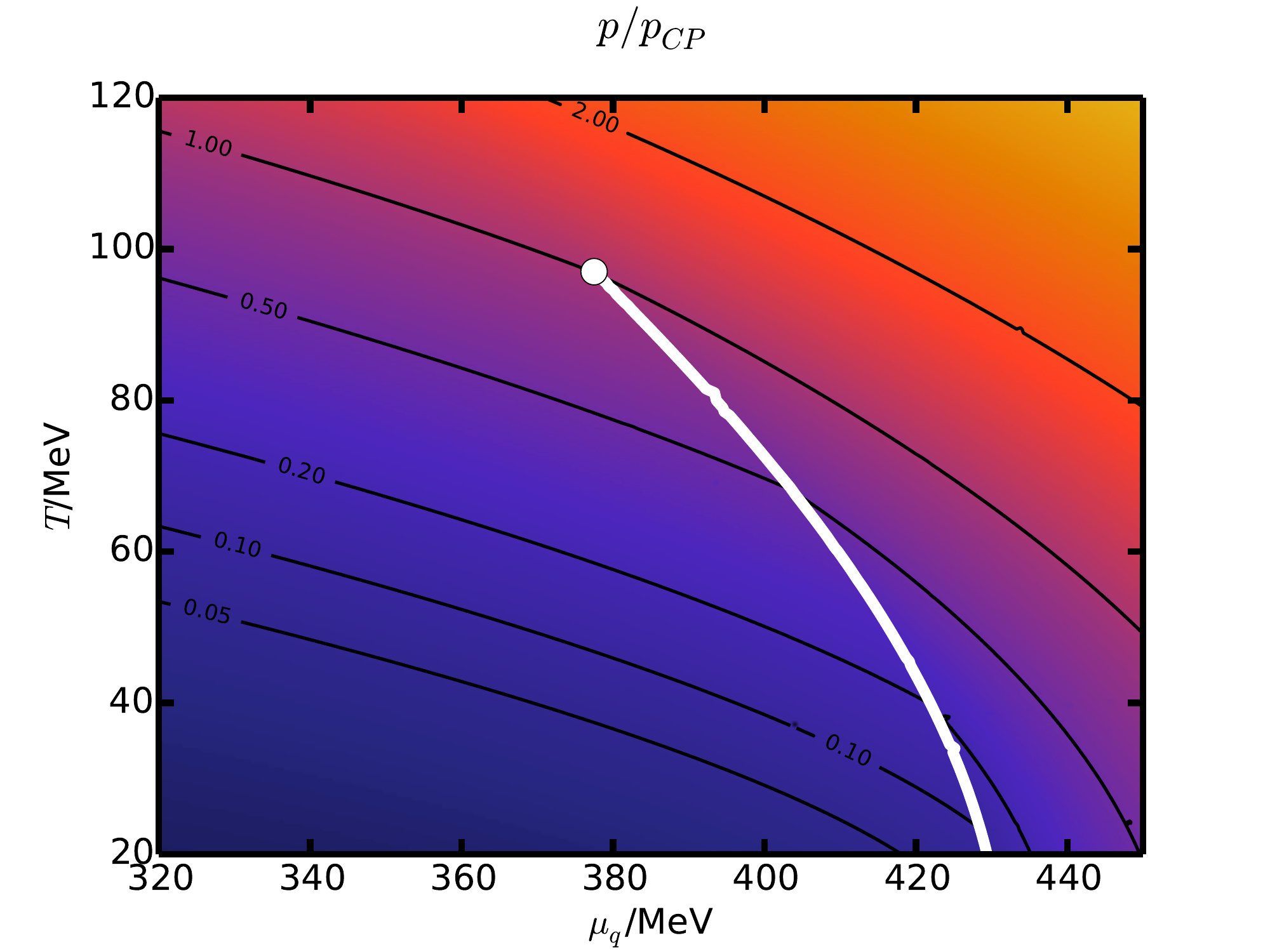}}
{\includegraphics[clip=true,trim=7mm 6mm 15mm 17mm,width=0.48\textwidth]{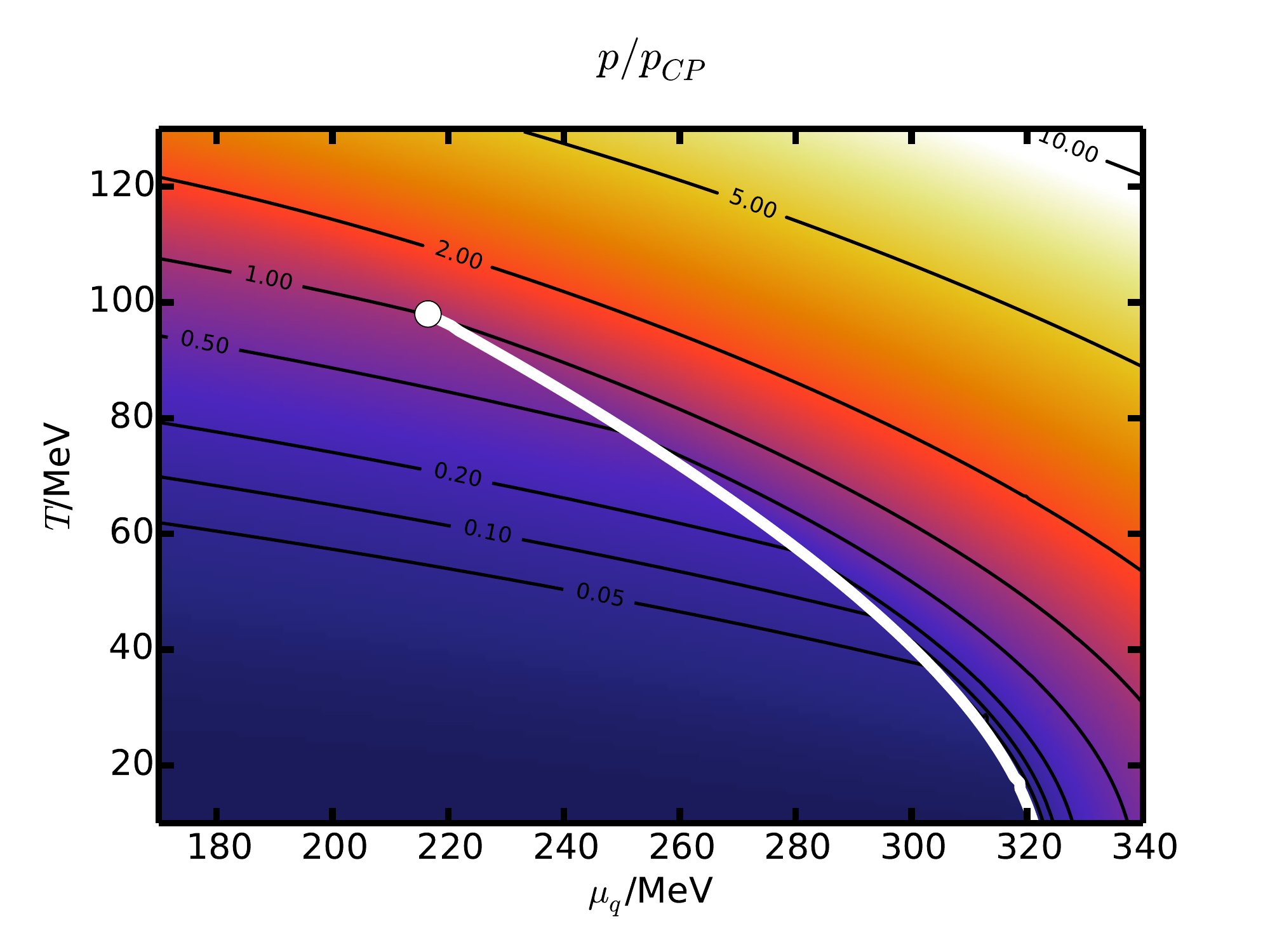}}\\
{\includegraphics[clip=true,trim=7mm 1mm 15mm 13mm,width=0.48\textwidth]{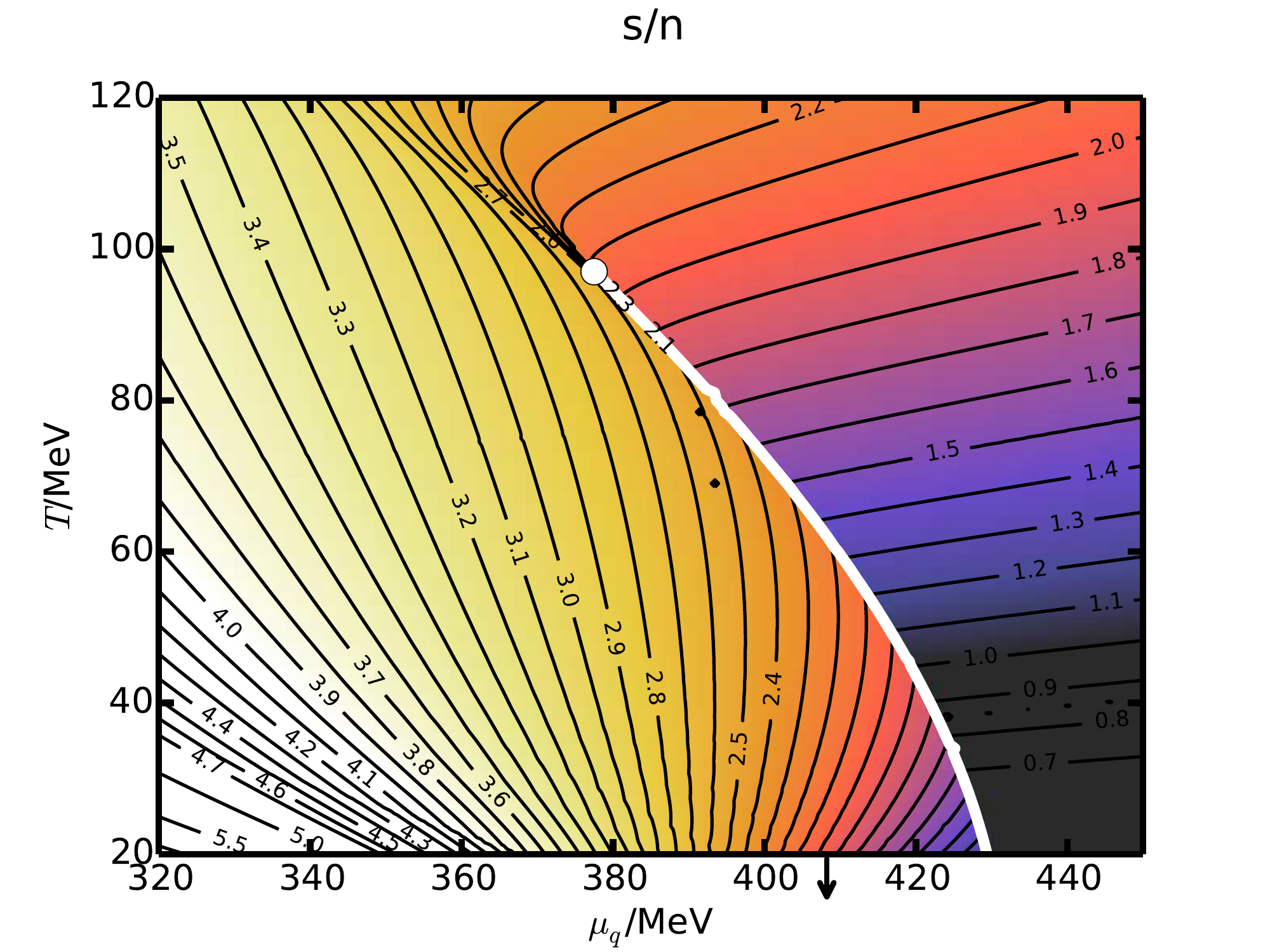}}
{\includegraphics[clip=true,trim=7mm 6mm 15mm 17mm,width=0.48\textwidth]{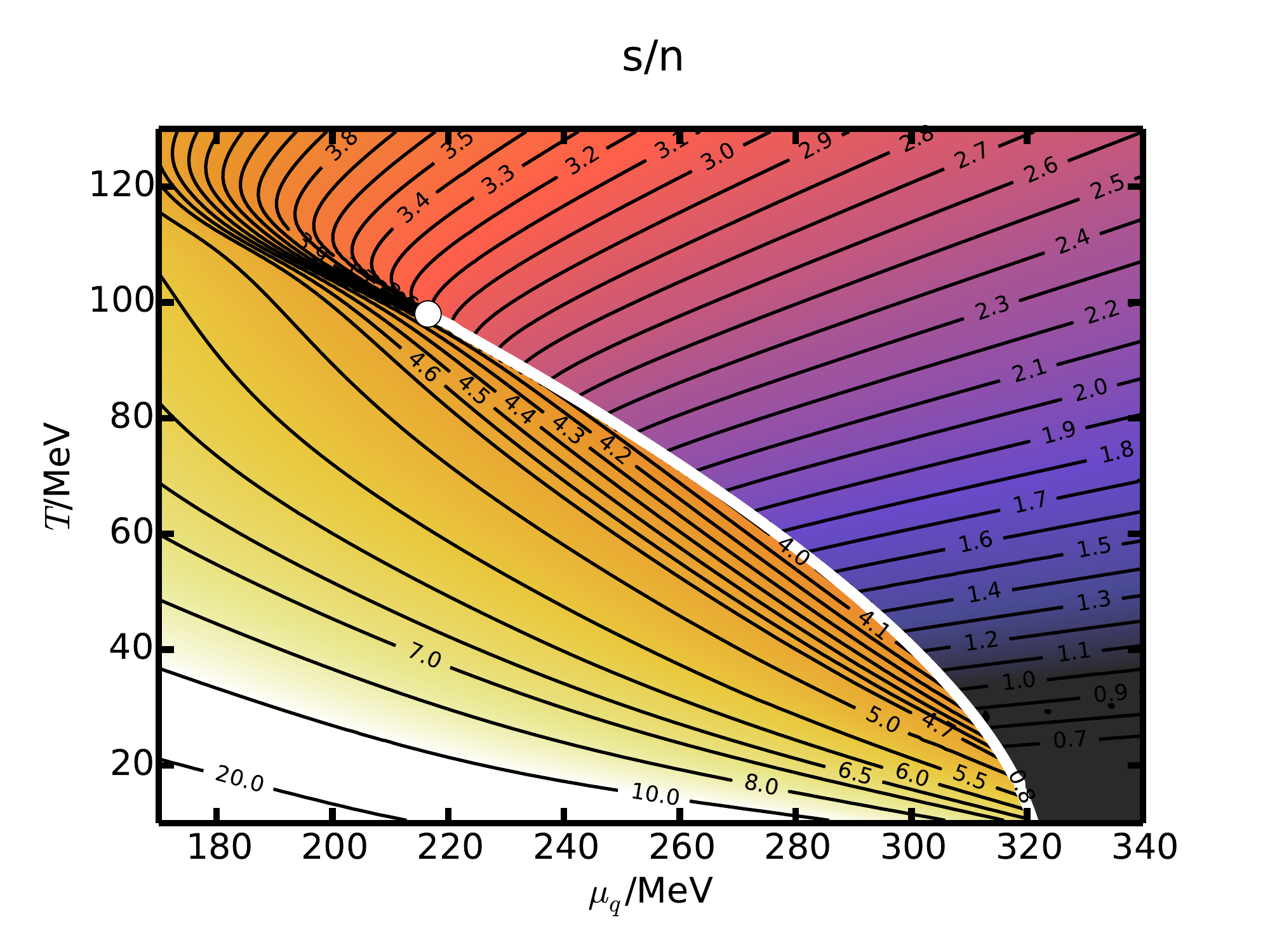}}
\caption{Contour plots of scaled pressure $p/p_{CEP}$
(i.e.\ isobars, top row) and
entropy per baryon $s/n$ (i.e.\ isentropes, bottom row) for FOPTs of 
type IA (left column) and type II (right column) over the \mbox{$T$--$\mu$} plane.
Equation of state from the quark-meson model with linearized fluctuations
applying the parameters $\langle \sigma \rangle_0 = 90$~MeV (expectation value 
of the sigma field in vacuum, as indicated by the label $0$),
$m_{\pi,0} = 138$~MeV (pion mass)
as well as either $m_{\sigma,0} = 1284.4$~MeV (sigma mass), $m_{q,0} = 390$~MeV (quark mass) (left column)
or  $m_{\sigma,0} = 700$~MeV, $m_{q,0} = 360$~MeV (right column). The pressure is scaled by the
pressure at the critical end point, i.e. with $p_{CEP}=2.38\times10^8 {\rm MeV^4}$ (left) and 
$p_{CEP}=8.59\times10^8 {\rm MeV^4}$ (right), respectively. The arrow in the bottom left plot points
to a state where the density at $T=0$ is equal to $n_0=0.17\,\text{fm}^{-3}$. On the bottom right plot this
point is located at the phase boundary.
\label{fig:1}}
\end{figure}

Figure \ref{fig:1} exhibits the isobars $p = const$ over the \mbox{$T$--$\mu$} plane
for two parameter sets (see figure caption for the values)
of the quark-meson model in linearized fluctuations approximation \cite{Mocsy:2004ab,Bowman:2008kc,Ferroni:2010ct,Wunderlich:2015rwa}.
These patterns look fairly similar at a first glance. The isobars are flatter than
the phase border line (fat white curve). The CEP coordinates are
$(T_{CEP}, \mu_{CEP}) = ({\cal O}(97 {\rm MeV}), {\cal O} (377.5 {\rm MeV}))$ for the parameter set
depicted on the left panels and $({\cal O}(98 {\rm MeV}), {\cal O} (216.5 {\rm MeV}))$ on the right ones.
(Note that we use actually quark chemical potential $\mu_q$ and net 
quark density $n_q$.)
One must not consider these values as predictions of the CEP location since the proper
account of fluctuations can significantly  change them. Furthermore, the inclusion
of some gluon dynamics, e.g.\ via a coupling to the Polyakov loop, thermal gluon
fluctuations as well as extending the invoked hadron species can also cause substantial
changes of the CEP coordinates.

Despite of the apparently marginal differences of the isobar patterns, the isentropes 
are drastically different. In the left bottom panel of Fig.~\ref{fig:1}, type IA 
isentropes are seen
which mean incoming from phase "2" and outgoing into phase "1" whenever they
meet the critical curve. In contrast, the right bottom panel in Fig.~\ref{fig:1}
displays a type II FOPT with incoming-only isentropes into the critical curve.
      
\begin{figure}
   {\includegraphics[clip=true,trim=9mm 2mm 17mm 15mm, width=0.48\textwidth]{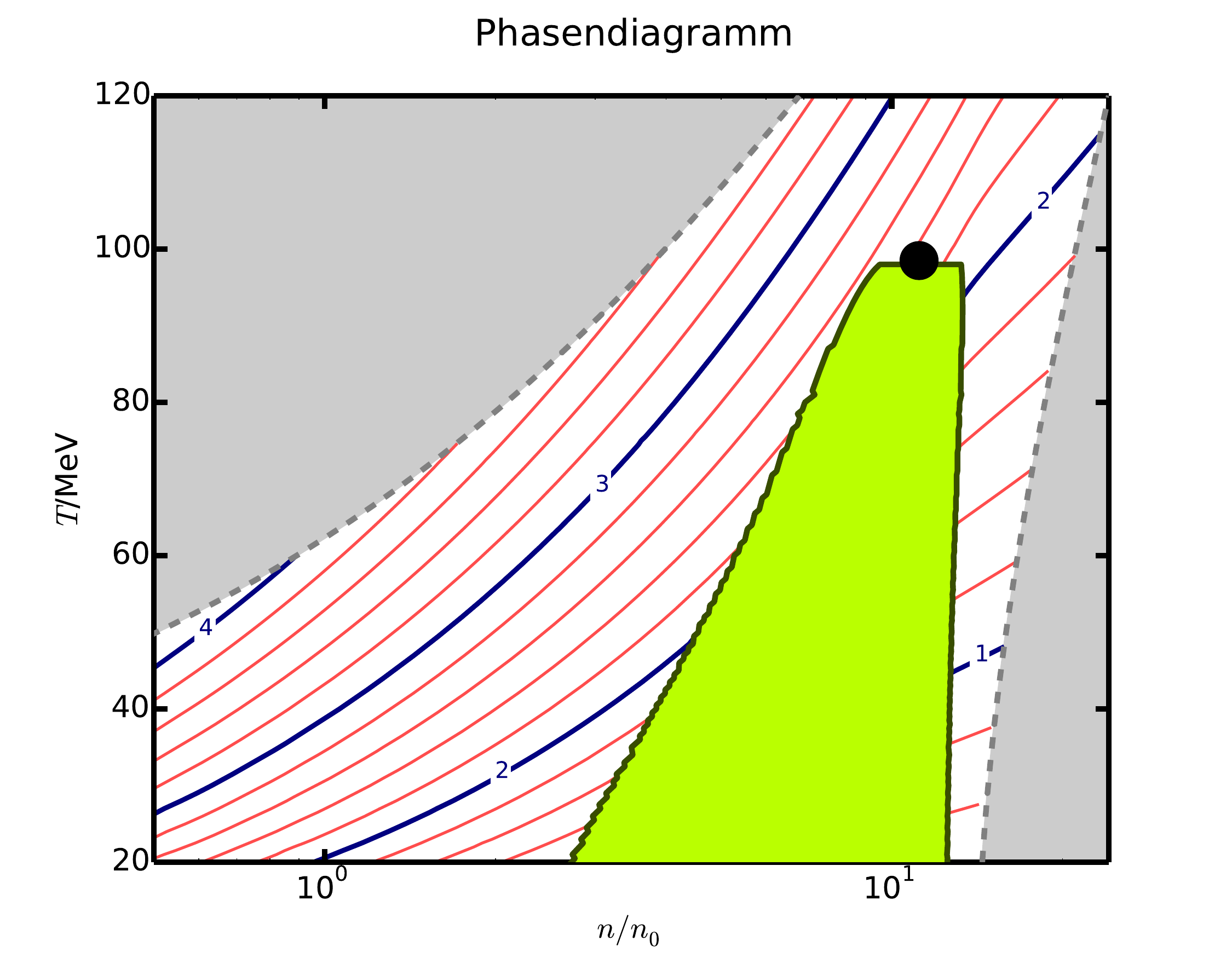}}
   {\includegraphics[clip=true,trim=9mm 2mm 17mm 15mm, width=0.48\textwidth]{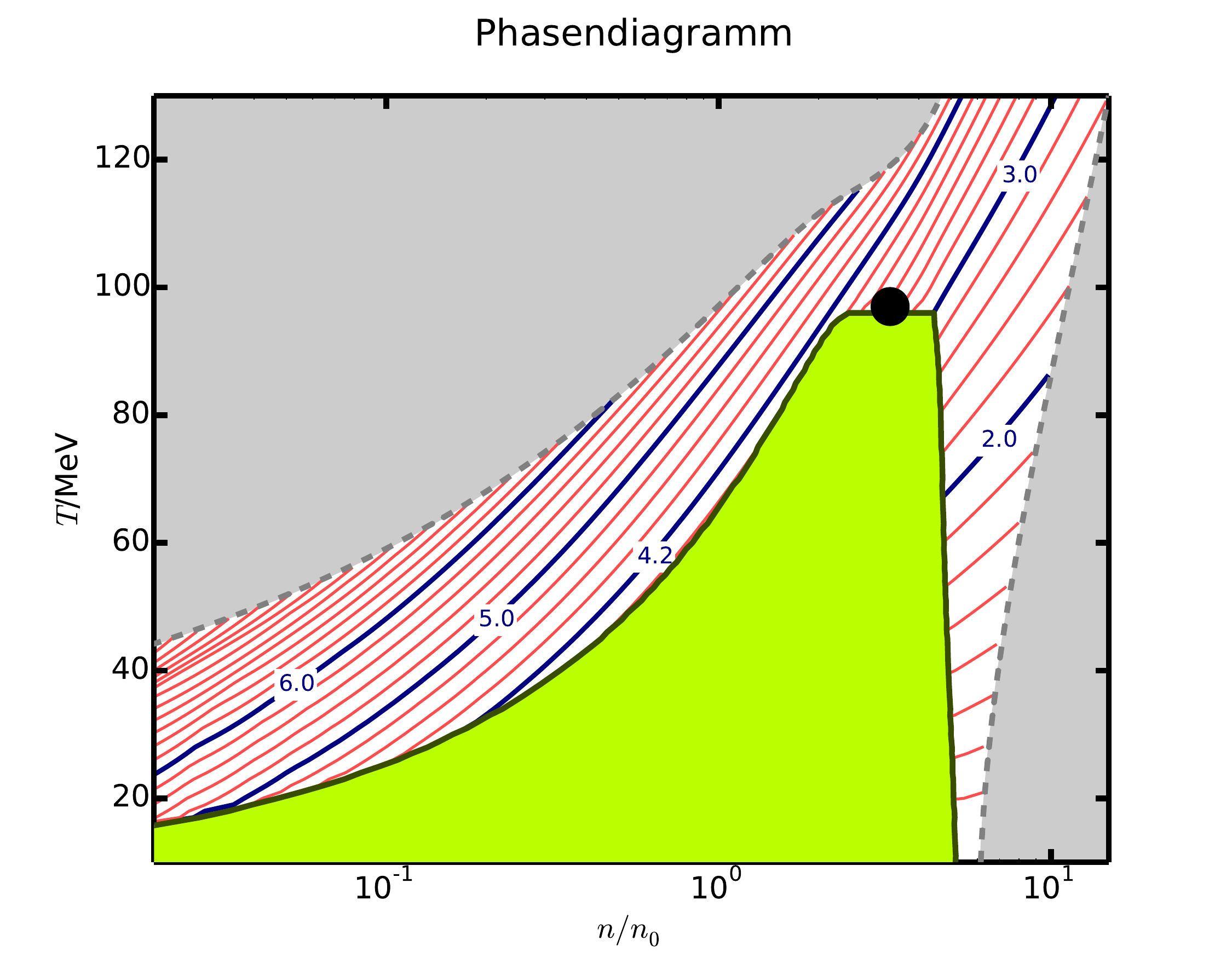}}
   \caption{As Fig.~{\ref{fig:1}} but for the isentropes in the \mbox{$T$--$n$} plane
            for pure phases only. The difference in $s/n$ between two adjacent isentropes is 0.2 and the thick blue 
            isentropes are labeled with their respective $s/n$.
            The two-phase coexistence regions are depicted as 
            green areas with the CEP (black bullet) on top. The dashed grey curves enclose the regions 
            in \mbox{$T$--$\mu$} space displayed in Fig.~\ref{fig:1}, i.e. the gray regions correspond to regions outside.
            The densities are scaled by the nuclear saturation density
            $n_0 = 0.17\,\rm fm^{-3}$. 
   \label{fig:2}}
\end{figure}

Figure \ref{fig:2} exhibits the isentropes in pure phases "2" and "1" over the
\mbox{$T$--$n$} plane. This presentation verifies that both the temperature and the 
density drop along the isentropes in pure phases. One can infer directly from
the bottom panels of Fig.~\ref{fig:1} the above claim w.r.t.\ outgoing isentropes
from the low-density phase border curve $n_1 (T)$ for type IA, see left panel
of  Fig.~\ref{fig:2}, while for type II (right panel) only incoming isentropes 
appear (isentropes with $s/n>5$ enter the two-phase region at smaller densities which are not displayed). 

\begin{figure}
\includegraphics[width=0.48\textwidth]{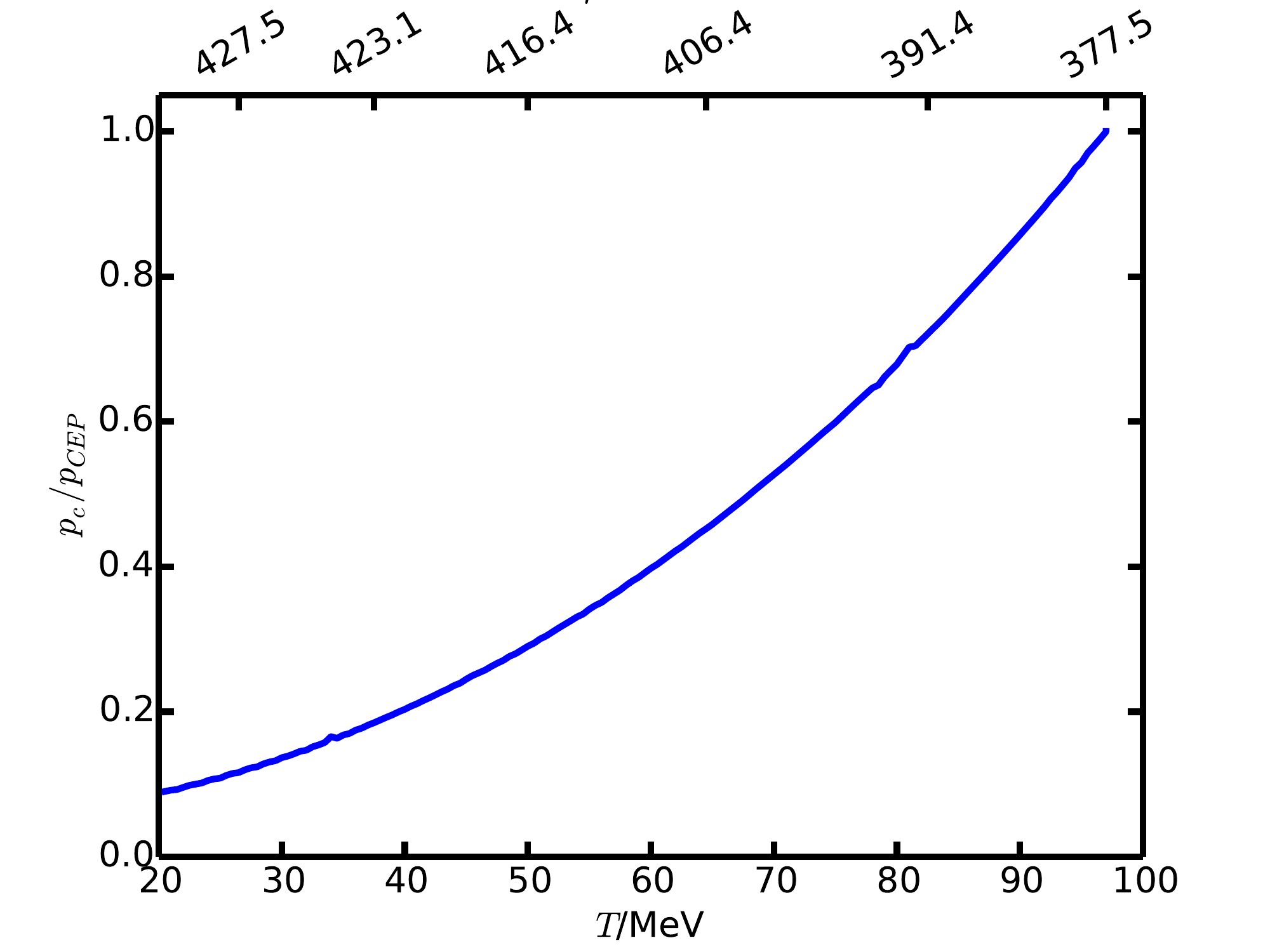}
\includegraphics[width=0.48\textwidth]{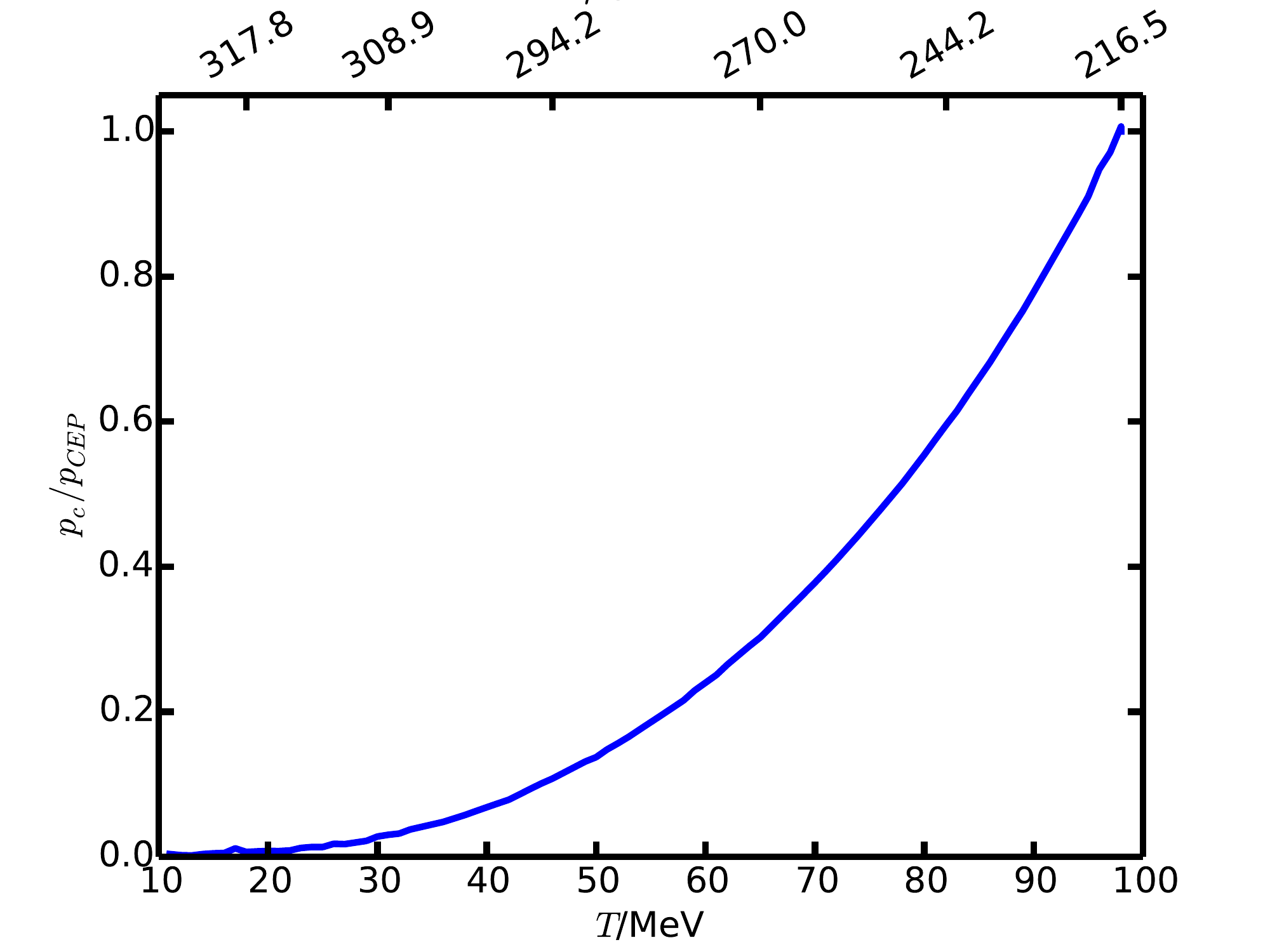}
\caption{The critical pressure $p_c(T)$ as a function of temperature
for FOPTs of type IA (left panel) and II (right panel). 
The numbers on the upper axis are the critical chemical potentials (in MeV) corresponding to the 
temperatures on the lower axis.
Equation of state and critical pressures $p_{CEP}$ as
described in the caption of  Fig.~{\ref{fig:1}}.
\label{fig:3}}
\end{figure}

Consistent to the Clausius-Clapeyron equation \eqref{clausius_1}, the critical pressure
as a function of the temperature is increasing, see Fig.~\ref{fig:3}. The inclined numbers at the 
top axis depict the (critical) chemical potential values corresponding to the temperature given at the lower axis
thus highlighting the shape of $p_c(\mu)$ which is actually decreasing in agreement with \eqref{clausius_2}.

We mention that the employed minimum set-up of the quark-meson model does not allow for 
type IB transitions since thermal gluon fluctuations are not included, i.e. the number of 
effective degrees of freedom accounting for thermal fluctuations is too small. One may,
however, easily construct two-phase models with a high-temperature quark-gluon
phase and a low-temperature hadron phase. Figure \ref{fig:4} in the appendix \ref{apdx:HQ} presents such an
example. 
Without fine tuning, such models do not display a CEP at $\mu > 0$, instead 
the constructed phase border curve continues form the $T$ axis down to the
$\mu$ axis. Reference \cite{Steinheimer:2012gc} provides an example of 
enforcing a CEP at $\mu > 0$ to obtain also a type IB transition.

The focus of the present note is on the isentropes relevant for the expansion dynamics in
relativistic heavy-ion collisions. As emphasized, e.g. in \cite{Hempel:2015vlg} and references therein,
analog considerations are useful for discussing the impact of peculiarities of the 
QCD phase diagram in core-collapse supernova explosions.
There, one has to consider adiabatic paths along compression with proper leptonic contributions
including also trapped neutrinos.
Tor a first orientation, the pressure as a function of the energy density at suitable values of the entropy per baryon 
is to be analyzed to figure out whether the FOPT effects in iso-spin symmetric matter translate 
into modifications of neutron star configurations (with $\beta$ stability, no trapped neutrinos)
such as the occurrence of a third stable island (cf. 
\cite{Kampfer_neutronstars}), 
nowadays often refered to as 
twin configurations \cite{Schertler:1997za,Schertler:1999xn,Schertler:2000xq,Alford:2013aca,
Drago:2015dea,Zdunik:2012dj,Chamel:2013efa,Alvarez-Castillo:2014dva,Blaschke:2013ana,Benic:2014jia},
or modify the core collapse dynamics (with trapped neutrinos) toward proto-neutron stars or even black holes such
as discussed in \cite{Hempel:2015vlg,Sagert:2008ka} and references therein. We leave according 
investigations to separate dedicated analyses.
}

\sectionn{Conclusions and Summary}
\label{sec:summary}

{ \fontfamily{cmr}\selectfont
 \noindent
In summary we discuss options for modeling a hypothetical first-order phase transition
which is related to a critical end point in a strongly interacting medium. 
Guided by the expectation that the QCD cross-over (as remnant of the transition of massless 2+1 flavor QCD,
cf. \cite{Ding:2015ona}) at a temperature of about 150~MeV at
small chemical potential turns, at the critical point at large chemical potential, into
a first-order transition we consider scenarios where initially deconfined matter can evolve completely into
confined (hadronic) matter. We emphasize that both enthalpic and entropic phase transitions
are consistent with such an expectation provided a graceful exit from the deconfined state
into pure hadron matter is possible upon adiabatic expansion. 
At low temperature, the low density part of the two-phase coexistence region must be at larger densities
than nuclear  matter at saturation (for isospin symmetric nuclear matter).
This implies that the pattern of isentropes must "go through"
the phase border curve to be conform with the envisaged scenario. In contrast, the 
van der Waals type transition is of a different kind as it has locally incoming isentropes only.
Obviously, more complicated phase border curves may allow for mixtures of the mentioned
types. 
Our discussion also completely ignores flavor-locked color superconducting phases which are expected
at larger densities.

Our discussion is based on equilibrium thermodynamics, and the medium is assumed to obey
one conserved charge - the baryon density. Accounting for more conserved charges, e.g.\
related to isospin, strangeness, electric  charge etc., complicates the picture. Transient states
related to under saturated or over saturated gluons \cite{Peshier:2015kco} or under saturated quark
state occupation \cite{Stoecker:2015zea} give rise to many interesting phenomena beyond our
discussion. 

The lacking of ab intio information from first-principle calculations of QCD thermodynamics
lets many options still be conceivable. This makes the concerted experimental hunt for signals of the
critical end point and the related first-order transition so important. }

\appendix
\sectionn{A two-phase model for type IB}
\label{apdx:HQ}

{ \fontfamily{cmr}\selectfont
 \noindent
\begin{figure}[!ht]
   \centering
   {\includegraphics[clip=true,trim=2mm 2mm 3mm 3mm,width=0.48\textwidth]{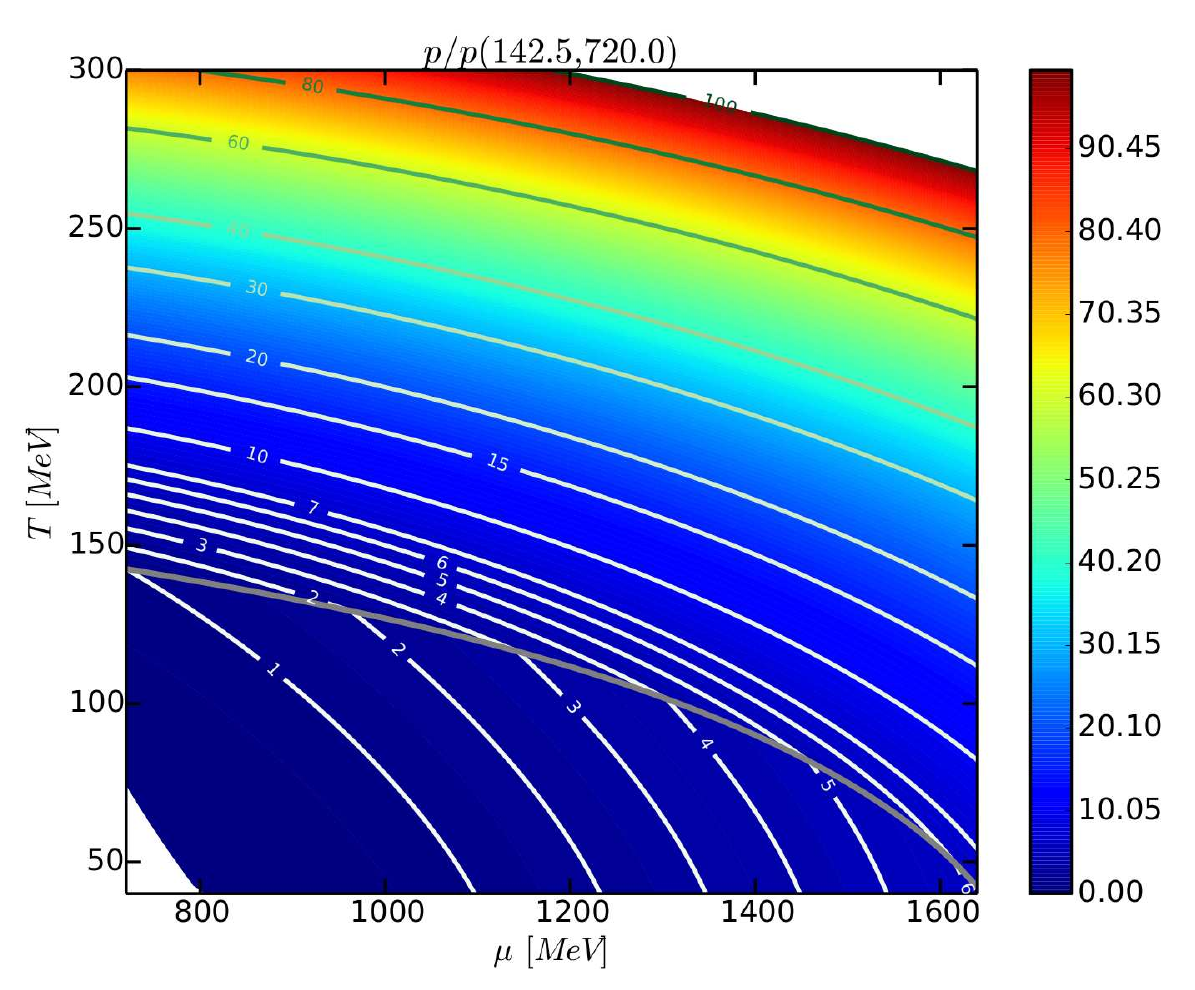}}\hfill
   {\includegraphics[clip=true,trim=2mm 4mm 5mm 5mm, keepaspectratio=false,width = 0.48\textwidth,height=0.395\textwidth]{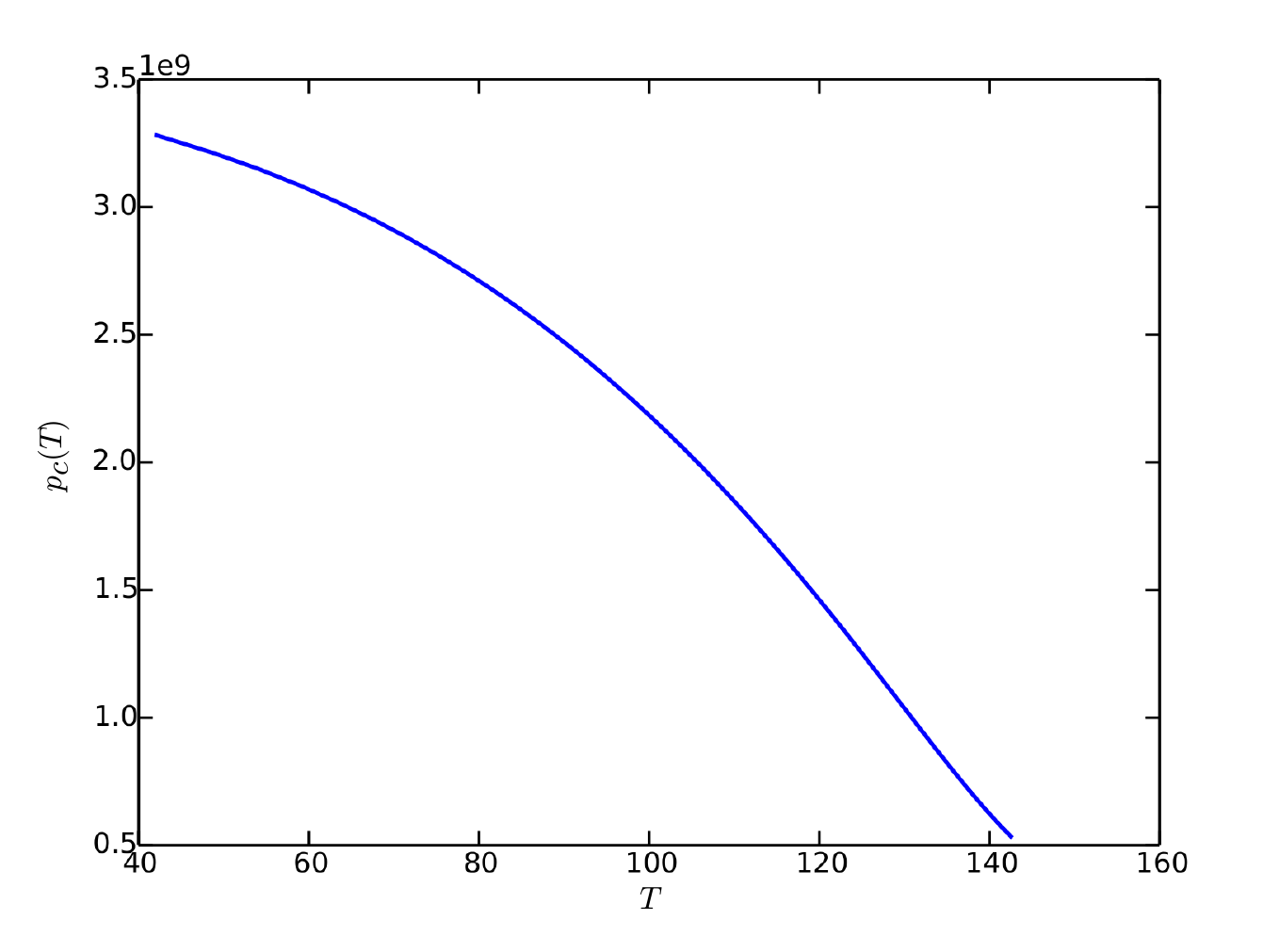}}\\
   {\includegraphics[clip=true,trim=4mm 2mm 3mm 3mm,width=0.48\textwidth]{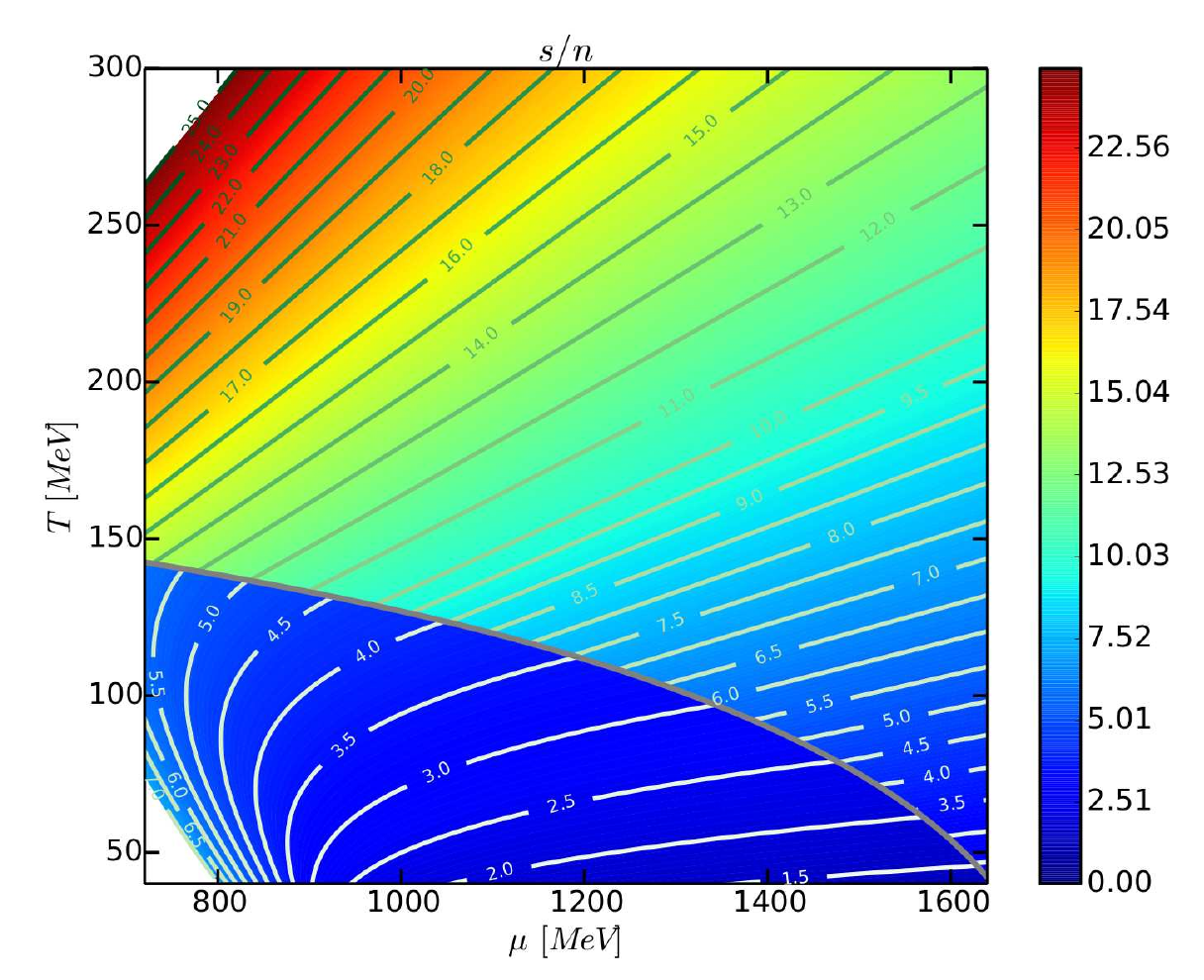}}\hfill
   {\includegraphics[clip=true,trim=4mm 4mm 1mm 1mm,width=0.48\textwidth]{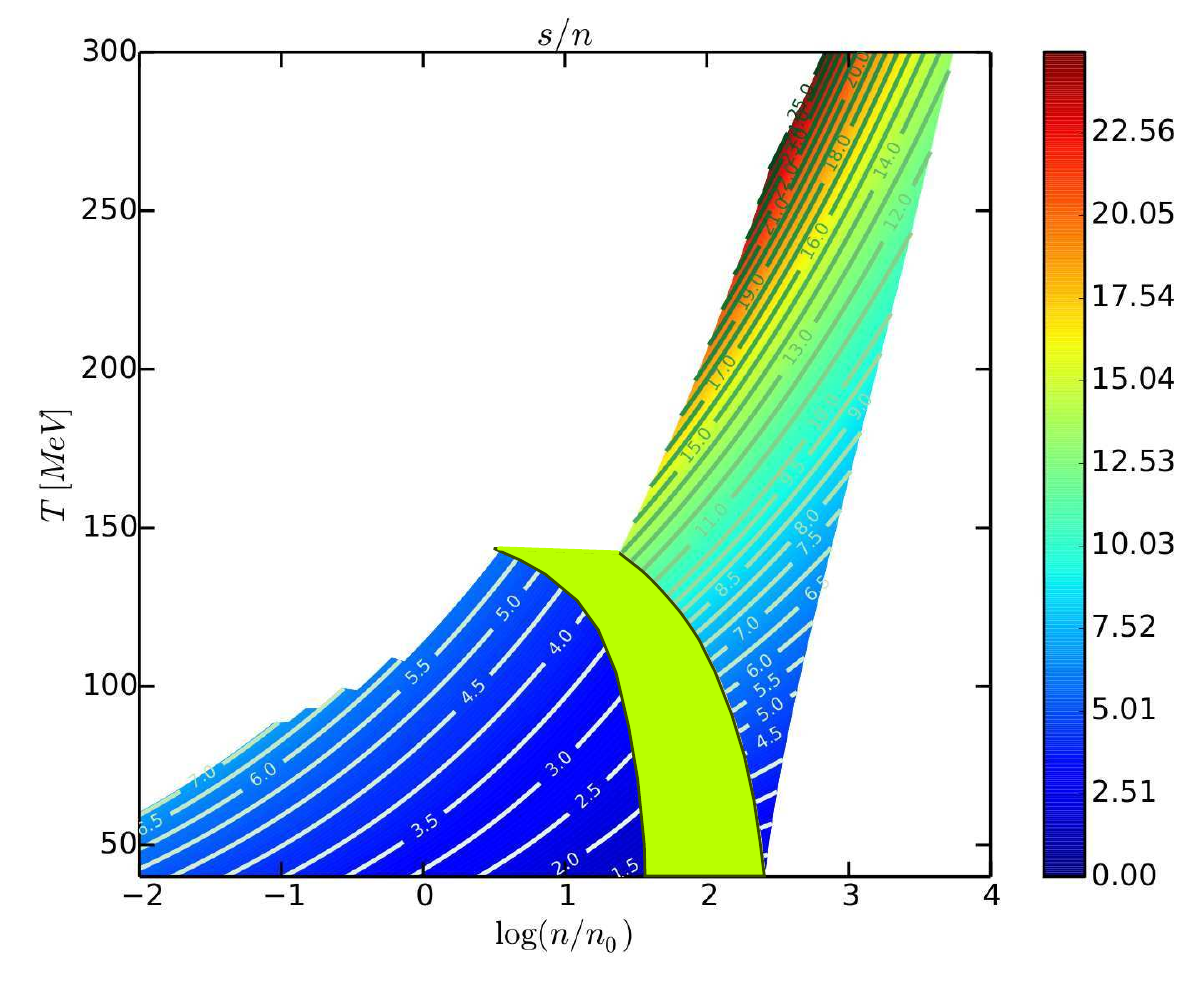}}
   \caption{Isobars (left top panel) and the critical pressure $p_c$ as a function of temperature (right top panel)
   as well as isentropes, both over the \mbox{$T$--$\mu$} plane (left bottom) and 
   over the \mbox{$T$--$n$} plane (right bottom) for the two-phase model of type IB FOPT, based on 
   \mbox{Eqs.~(\ref{HQ_01}-\ref{HQ_05})}. As in Fig.~\ref{fig:2}, the coexistence region is depicted as green area.
   Our calculations do not map out completely the \mbox{$T$--$\mu$} plane, thus leaving some uncharted regions
   in white in the left column and the bottom right panel.
   }
   \label{fig:4}
\end{figure}
The constructed FOPT is based on the extrapolation of a hadron equation of state with pressure
\begin{equation}
   p_1 = n_b^2 \frac{\partial W}{\partial n_b}\Bigg|_{\hat s}\label{HQ_01}
\end{equation}
to be calculated from 
\begin{equation}
   W = e/n_B = W_c + W_T \label{HQ_02}
\end{equation}
with
\begin{align}
   W_c        =&\frac{K}{18}\left(\frac{n_B-n_0}{n_B}\right)^2 + W_\text{bind} + m_N,\notag\\
   W_T        =&\frac32 T + \frac{\pi^2}{18}\frac{T^4}{n_B},\notag\\
   \hat s     =& \frac{s}{n_B} = \hat s_N + \hat s_\pi,\label{HQ_03}\\
   \hat s_N   =& 2.5-\ln\left(\frac{n_B}{4}\left(\frac{2\pi}{Tm_N}\right)^{3/2}\right),\notag\\
   \hat s_\pi =& \frac{4}{3}\frac{\pi^2}{10}\frac{T^3}{n_B}.\notag
\end{align}
The temperature $T(n_B,\hat s)$ follows self consistently from 
\begin{equation}
   T=\frac{\partial W(n_B, \hat s)}{\partial \hat s}\label{HQ_04}
\end{equation}
and the baryo-chemical potential is then $\mu_B = W + p/n_B - T\hat s$. We utilize the nucleon mass $m_N=938$~MeV, 
the nucleon binding energy \mbox{$W_\text{bind}=-16$~MeV}, nuclear incompressibility coefficient 
$K= 235$~MeV and saturation density $n_0=0.17$~fm$^{-3}$.\footnote{
This is a model in the spirit of \cite{Buchwald:1981hu} for nuclear matter and pions.}
The equation of state in the high temperature phase is defined by the extrapolation of a quark-gluon equation of state 
from leading-order weak-coupling (cf. \cite{Kurkela:2016was} for advanced calculations) supplemented by a bag constant B
\begin{equation}
   p_2 = 16\frac{\pi^2}{90}T^4 + f_q\Big(\frac78\frac{\pi^2}{90}T^4 +\frac{1}{24}T^2\mu_B^2 + \frac{1}{48\pi^2}\mu_B^4\Big)-B,\label{HQ_05}
\end{equation}
where we employ for the number of effective quark degrees of freedom $f_q=2.5\times3\times2\times2=30$ and $B = (235$~MeV$)^4$. 
These branches are matched by the above mentioned Gibbs criteria for equilibrium, $p_1=p_2$, $T_1=T_2$, $\mu_1 = \mu_2$.
The resulting isobars, the critical pressure $p_c(T)$ as well as isentropes, both over the \mbox{$T$--$\mu$} and the 
\mbox{$T$--$n$}-planes
are exhibited in Fig.~\ref{fig:4}.

}

 {\color{myaqua}

 \vskip 6mm

 \noindent\Large\bf Acknowledgments}

 \vskip 3mm

{ \fontfamily{cmr}\selectfont
 \noindent
 We tank J. Randrup, V. Koch, F. Karsch, K. Redlich, M.I. Gorenstein, S. Schramm, H. St\"ocker 
and B. Friman for enlightening
discussions of phase transitions in nuclear matter. The work is supported by BMBF grant 05P12CRGH. 
}

\bibliographystyle{aip_jmp}
\bibliography{FWunderlich_2016}
\end{document}